\documentclass[useAMS,usenatbib,letters]{mnras}

\usepackage{graphicx, bm, amssymb, xcolor}
\usepackage{verbatim}
\usepackage[all]{hypcap}
\usepackage{xspace}
\usepackage{multirow,bigdelim}
\usepackage[fleqn]{amsmath}

\usepackage{epsfig}
\usepackage{aas_macros}
\usepackage{natbib}
\usepackage{times,txfonts}
\usepackage{morefloats}
\usepackage{tabularx}
\usepackage{lscape}

\newcommand{\msun}{\mbox{M$_\odot$}}

\newcommand{\gyr}{\mbox{${\rm Gyr}$}}

\newcommand{\dex}{\mbox{${\rm dex}$}}
\newcommand{\feh}{\mbox{$[{\rm Fe}/{\rm H}]$}}
\newcommand{\oh}{\mbox{$12+\log({\rm O}/{\rm H})$}}
\newcommand{\zsun}{\mbox{Z$_\odot$}}
\newcommand{\zstar}{\mbox{$Z_\star$}}
\newcommand{\zgas}{\mbox{$Z_{\rm gas}$}}
\newcommand{\mi}{\mbox{$M_{\rm i}$}}
\newcommand{\mh}{\mbox{$M_{\rm h}$}}
\newcommand{\ms}{\mbox{$M_\star$}}

\newcommand{\be}{\begin{equation}}
\newcommand{\ee}{\end{equation}}
\newcommand{\bea}{\begin{eqnarray}}
\newcommand{\eea}{\end{eqnarray}}

\newcommand{\appropto}{\mathrel{\vcenter{
  \offinterlineskip\halign{\hfil$##$\cr
    \propto\cr\noalign{\kern1pt}\sim\cr\noalign{\kern-2pt}}}}}

\title[Minimum metallicity of globular clusters]{\vspace{-6mm}The minimum metallicity of globular clusters and its physical origin -- implications for the galaxy mass-metallicity relation and observations of proto-globular clusters at high redshift\vspace{-5mm}}

\author{J.~M.~Diederik Kruijssen\thanks{E-mail: \href{kruijssen@uni-heidelberg.de}{kruijssen@uni-heidelberg.de}}\\
Astronomisches Rechen-Institut, Zentrum f\"{u}r Astronomie der Universit\"{a}t Heidelberg, M\"{o}nchhofstra\ss e 12-14, 69120 Heidelberg, Germany
\vspace{-4mm}}

\begin{document}

\date{Accepted 2019 April 10. Received 2019 April 9; in original form 2019 March 17\vspace{-3mm}}

\pagerange{\pageref{firstpage}--\pageref{lastpage}} \pubyear{2019}

\maketitle

\label{firstpage}

\begin{abstract}
In the local Universe, globular clusters (GCs) with metallicities $\feh<-2.5$ are extremely rare. In this Letter, the close connection between GC formation and galaxy evolution is used to show that this GC metallicity `floor' results from the galaxy mass-metallicity relation of ultra low-luminosity galaxies (ULLGs) at high redshift, where the most metal-poor GCs must have formed. Galaxies with metallicities $\feh\la-2.5$ have too low masses to form GCs with initial masses $\mi\ga10^5~\msun$, needed to survive for a Hubble time. This translates the galaxy mass-metallicity relation into a maximum initial cluster mass-metallicity relation for $\feh\la-1.8$, which naturally leads to the observed colour-magnitude relation of metal-poor GCs at $z=0$ (the `blue tilt'). Its strength traces the slope of the gas phase mass-metallicity relation of ULLGs. Based on the observed blue tilt of GCs in the Virgo and Fornax Clusters, the galaxy mass-metallicity relation is predicted to have a slope of $\alpha=0.4\pm0.1$ for $10^5\la\ms/\msun\la10^7$ at $z\ga2$. The GC metallicity floor implies a minimum host galaxy mass and a maximum redshift for GC formation. Any proto-GCs that may be detected at $z>9$ are most likely to end up in galaxies presently more massive than the Milky Way, whereas GCs in low-mass galaxies such as the Fornax dSph ($\ms\approx4\times10^7~\msun$) formed at $z\la3$.
\end{abstract}

\begin{keywords}
galaxies: evolution --- galaxies: formation --- galaxies: haloes --- galaxies: star formation --- globular clusters: general\vspace{-4mm}
\end{keywords}

\section{Introduction} \label{sec:intro}
Recent reports of relatively compact, young stellar populations in high-redshift ($z=2{-}8$) galaxies \citep[e.g.][]{bouwens17,johnson17,vanzella17,vanzella19} have caused considerable excitement in the globular cluster (GC) and galaxy formation communities. With the launch of the {\it James Webb Space Telescope} (JWST) creeping ever closer, the realisation has emerged that the formation of the progenitors of present-day GCs (often referred to as `proto-GCs') may be directly observable \citep{renzini17,boylankolchin18}, and that this highly clustered mode of star formation may even be dominant in ultra low-luminosity galaxies (ULLGs) forming in the early Universe \citep{zick18}. While it remains to be seen if the observed systems are real detections of proto-GCs or instead represent cluster complexes, as well as whether GC formation is truly expected to dominate star formation in ULLGs \citep{pfeffer19}, progress in this direction will be aided by quantitative predictions. Most importantly, at which redshifts and in which dark matter haloes do we expect to find the progenitors of present-day GCs? And if any are found, in which haloes are these GCs expected to end up at $z=0$?

In the local Universe, GC populations share a striking feature that may shed light on the above questions. Very few GCs have been found with metallicities $\feh<-2.5$ \citep[e.g.][]{forbes18,beasley19}. Because galaxies chemically enrich with time and follow a positive mass-metallicity relation \citep{maiolino19}, this implies the existence of a minimum galaxy mass and a maximum redshift for the formation of (proto-)GCs (also see \citealt{choksi18}). In this Letter, the galaxy mass-metallicity relation is used to explain the observed minimum GC metallicity around $\feh\approx-2.5$, to infer its implications for the origin of the `blue tilt' (the colour-magnitude relation of metal-poor GCs), which is shown to constrain the gas phase mass-metallicity relation of ULLGs ($10^5\la\ms/\msun\la10^7$) at $z\ga2$, and to place limits on the demographics of the proto-GC population, to be observed with current and future facilities (e.g.\ JWST) using gravitational lensing. \looseness=-1

\section{Physical origin of a GC metallicity floor}
Inferring in which galaxies the lowest-metallicity GCs may have formed requires knowledge of the galaxy mass-metallicity relation at high redshifts and low galaxy masses. This is exactly the regime where observational constraints are extremely scarce, reaching metallicities only as low as $\sim10^{-1}$ solar at $z\sim4$ \citep{maiolino19}. This is still more than an order of magnitude higher than the GC metallicity floor. However, high-resolution numerical simulations of galaxy formation predict the mass-metallicity relation as a function of redshift down to $10^{-3}$ solar. In this work, the mass-metallicity-redshift relations from the FIRE simulations are used \citep{ma16}, which are consistent with those from FIRE-2 \citep{hopkins18}. While this implies that the presented results are model-dependent, it is the only way of providing a meaningful interpretation of the GC metallicity floor in terms of its implications for galaxy evolution and the demographics of proto-GC host galaxies at high redshift.

\citet{ma16} provide the following mass-metallicity-redshift relations. For the stellar phase, they find
\be
\label{eq:zstar}
\log\left(\frac{\zstar}{\zsun}\right) = 0.40\left[\log\left(\frac{\ms}{\msun}\right) - 10\right] + 0.67 {\rm e}^{-0.50z} - 1.04 ,
\ee
and for the gas phase, they find
\be
\label{eq:zgas}
\log\left(\frac{\zgas}{\zsun}\right) = 0.35\left[\log\left(\frac{\ms}{\msun}\right) - 10\right] + 0.93 {\rm e}^{-0.43z} - 1.05 ,
\ee
which is converted to $\feh$ by writing $\feh=\log(Z/\zsun)-0.2$, as obtained by fitting the stellar metallicities \citep{ma16}. This conversion is necessarily an approximation, because it assumes constant chemical abundance ratios (consistent with observations at $\feh<-1.5$, see \citealt{tolstoy09}), but this is an acceptable uncertainty given the broad-brush strokes nature of this work.

\begin{figure}
\includegraphics[width=\hsize]{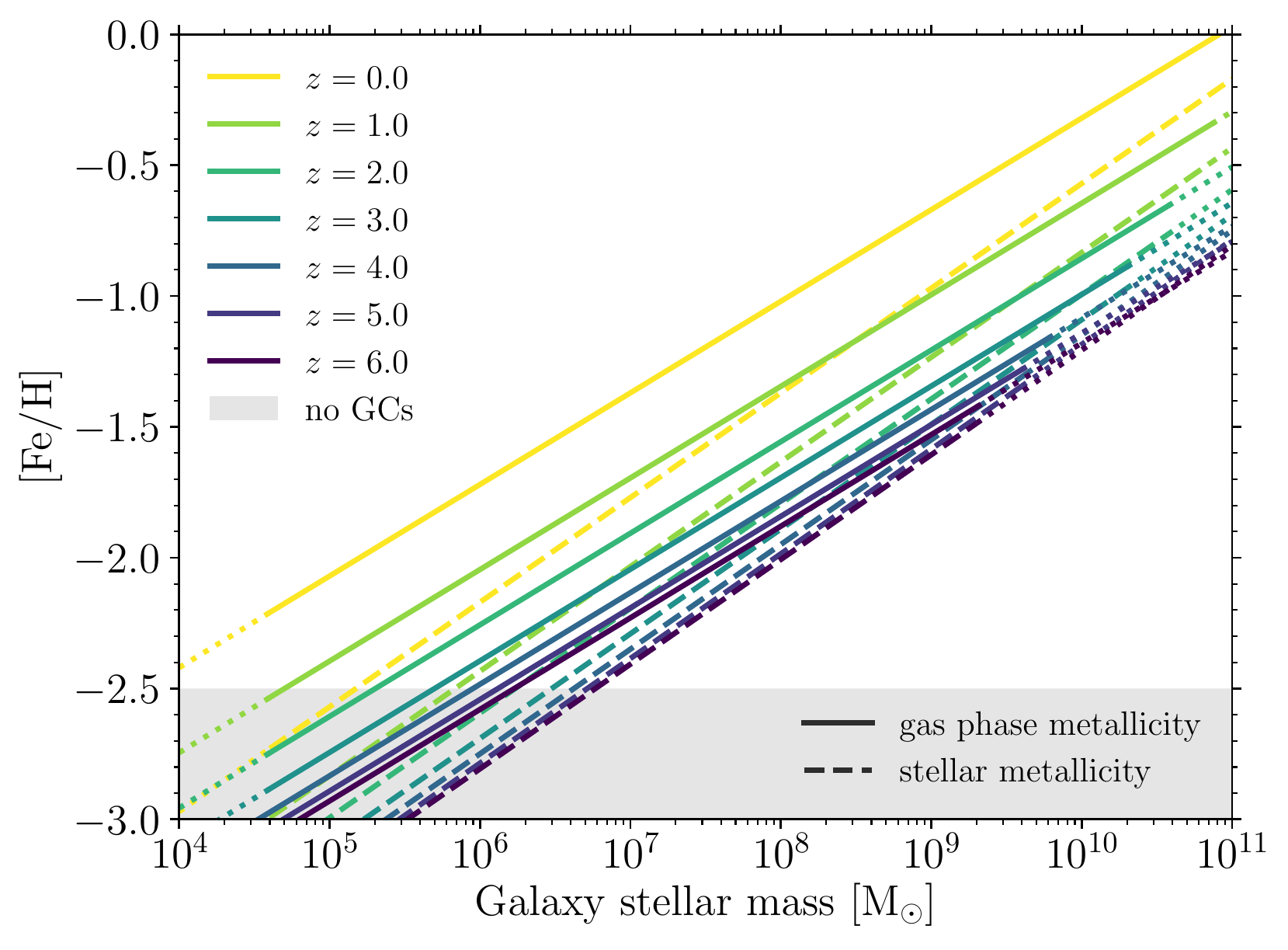}%
\caption{
\label{fig:f1}
Redshift evolution of the galaxy mass-metallicity relation from \citet{ma16} for the stellar phase (dashed lines) and the gas phase (solid lines) metallicity. The grey-shaded area indicates the metallicities at which few-to-no GCs are found at $z=0$. Throughout this work, dotted lines indicate extrapolations of the adopted galactic scaling relations. Galaxies at $z>2$ forming stars with $\feh=-2.5$ have stellar masses similar to those of GCs, suggesting that the paucity of GCs with $\feh<-2.5$ results from too low host galaxy masses.
}
\end{figure}
\autoref{fig:f1} visualises the above two relations. Both the stellar and gas phase relations are shown throughout this work, but the main focus is on the gas phase mass-metallicity relation, because it reflects the metallicity of stars and stellar clusters being formed at a specific redshift, rather than integrating over the mass formed previously. The lines in \autoref{fig:f1} clearly show that the mass-metallicity relations evolve by less than $0.3~\dex$ for $z>2$ and are thus largely independent of redshift for the epoch of cosmic history that this Letter focuses on, especially given that they exhibit a scatter of a similar magnitude (i.e.\ $0.3~\dex$) at $z>2$. The key insight drawn from \autoref{fig:f1} is that galaxies at $z>2$ forming stars with $\feh=-2.5$ have stellar masses of $\ms=10^5{-}10^6~\msun$, similar to the masses of GCs at $z=0$ \citep[e.g.][]{jordan07}. 

After their formation, clusters dynamically dissolve due to evaporation and tidal perturbations, leading to a minimum mass needed for their survival over a Hubble time. For GCs, this mass scale is predicted to be $\mi\sim10^5~\msun$ \citep{kruijssen15b,reinacampos18}. The masses of galaxies at $z>2$ forming stars with $\feh=-2.5$ barely exceed this mass scale. This suggests that a GC metallicity floor exists at $\feh\approx-2.5$, because lower-metallicity clusters formed in galaxies of such low masses that the resulting cluster masses were too low to enable their survival until the present day. To test this idea, the galaxy masses at which stars of a given metallicity are formed in \autoref{fig:f1} can be translated to a maximum initial cluster mass. Two cases are considered here. In the first, the maximum cluster mass equals the stellar mass of the host galaxy. In the second, this mass scale is decreased by a factor of $\Gamma$, which indicates the fraction of star formation occurring in bound stellar clusters \citep{bastian08,kruijssen12d} and accounts for the expected existence of an unbound young population of field stars. Observations of nearby galaxies and theory find $\Gamma=1{-}50$~per~cent \citep{adamo15b}. A fiducial value of $\Gamma=10$~per~cent is adopted here. Both cases assume the star-forming gas mass in the galaxy is similar to the stellar mass, as is found for galaxies down to $\ms=10^9~\msun$ at $z=2{-}4$ \citep{tacconi18}, and conservatively place all clustered star formation in a single massive GC.

\begin{figure}
\includegraphics[width=\hsize]{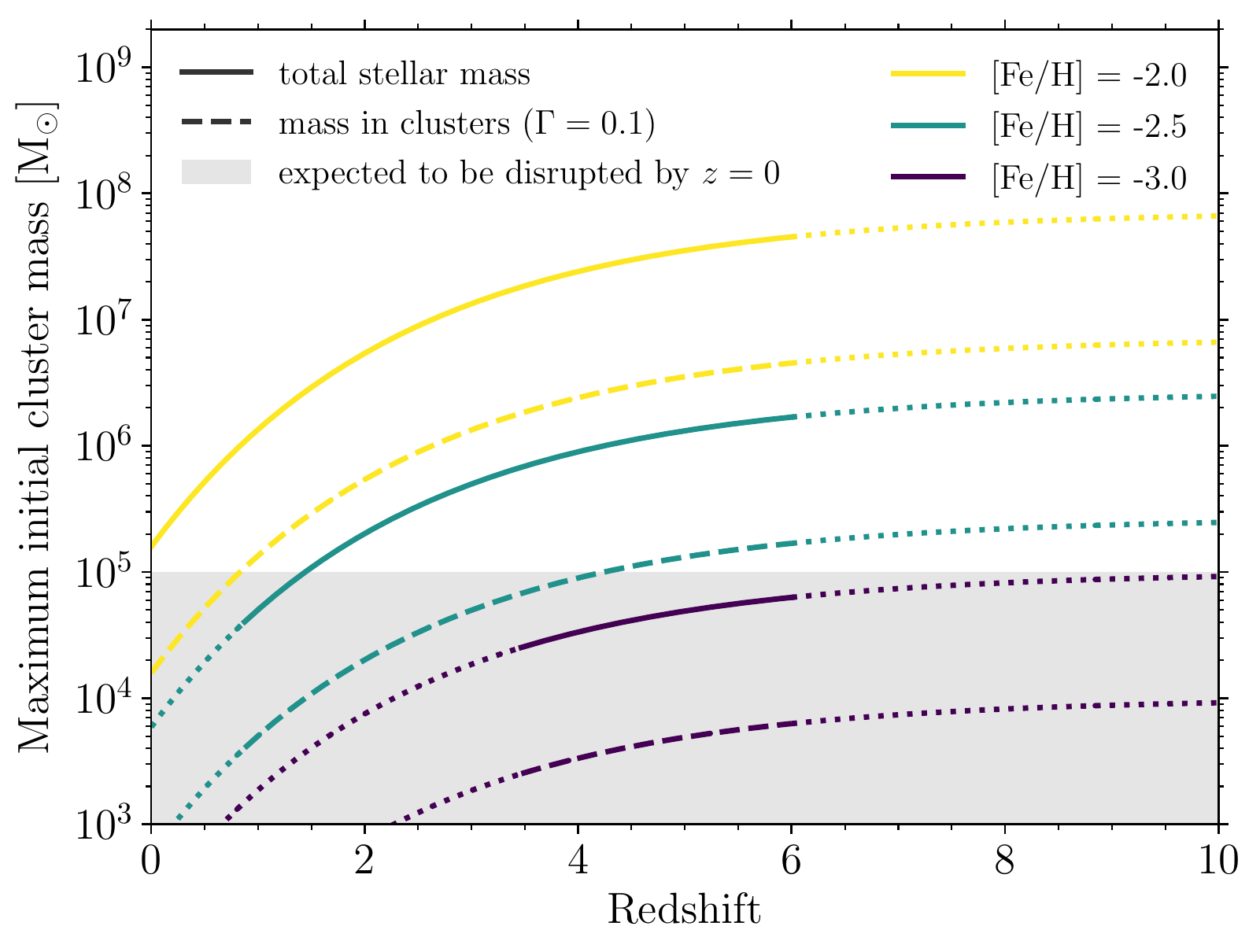}%
\caption{
\label{fig:f2}
Maximum mass of newborn proto-GCs as a function of redshift for different minimum GC metallicities ($\feh=\{-3.0,-2.5,-2.0\}$, top right legend). The maximum mass refers either to the total stellar mass of the host galaxy (solid lines) or 10~per~cent thereof (dashed lines), reflecting a more realistic situation in which not all stars form in bound stellar clusters (see the text). Clusters with initial masses $\mi\la10^5~\msun$ are unlikely to survive for a Hubble time (grey-shaded area), which applies to those with $\feh<-2.5$ and explains the GCs metallicity floor. 
}
\end{figure}
\autoref{fig:f2} predicts that the maximum mass of proto-GCs is a strong function of their metallicity. Clusters with $\feh<-2.5$ typically have masses $\mi<10^5~\msun$, which is likely too low to enable their long-term survival. This explains the minimum metallicity of GCs as a direct result of the galaxy mass-metallicity relation. The masses (and hence survival chances) of lower-metallicity clusters were limited by the low masses of their host galaxies.

\section{Implications for the mass-metallicity relations of GCs and ULLGs}
\begin{figure*}
\includegraphics[width=0.875\hsize]{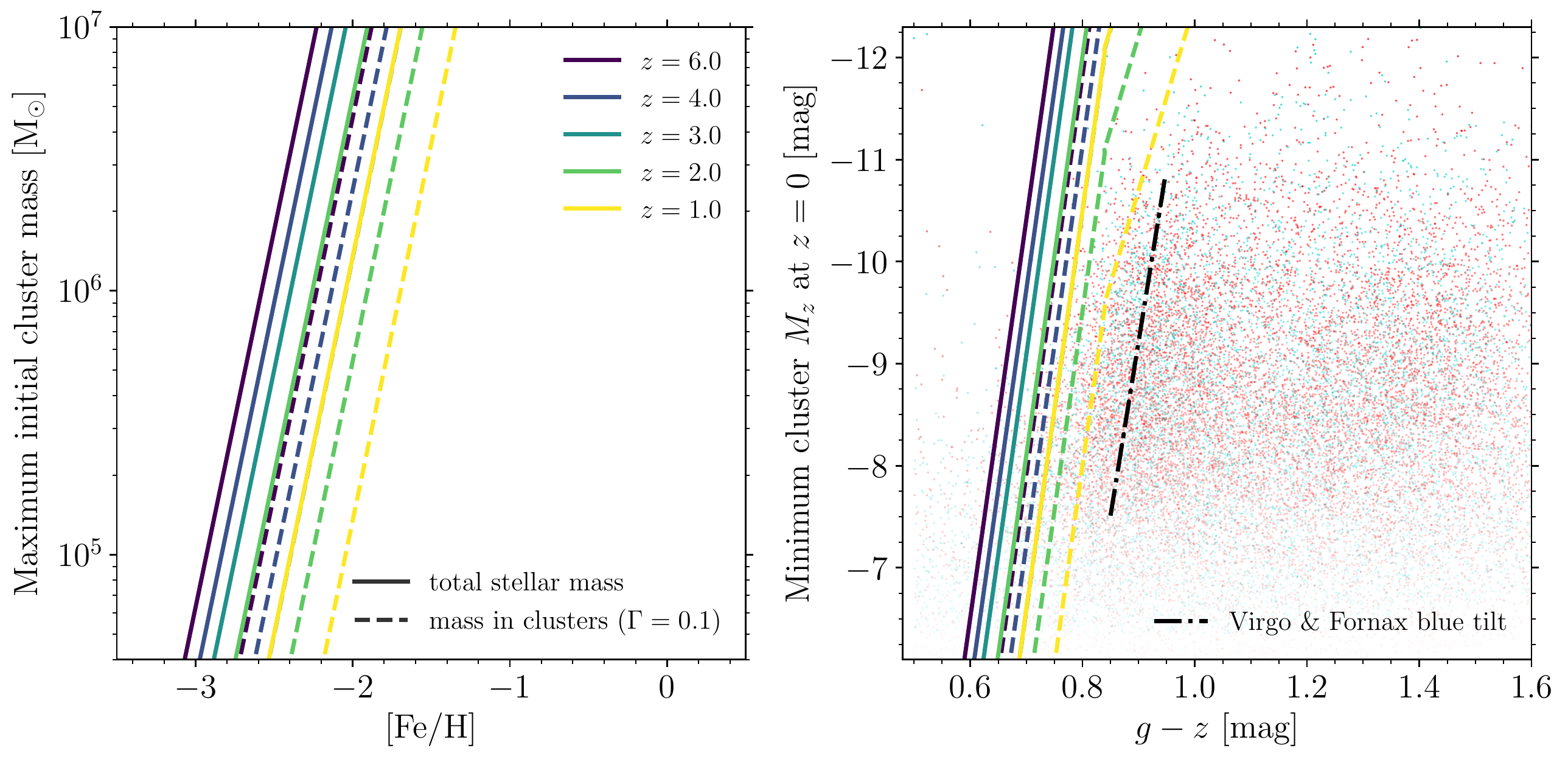}%
\caption{
\label{fig:f3}
The galaxy mass-metallicity relation sets the mass-metallicity (or colour-magnitude) relation of metal-poor GCs (i.e.~the `blue tilt'). Left panel: maximum initial cluster mass as a function of metallicity for different redshifts, for an available mass equal to that of the host galaxy (solid lines), or a lower mass to accommodate unbound field star formation (dashed lines). Right panel: colour-magnitude distribution of GCs in the Virgo (red) and Fornax (cyan) Clusters (data points). The dash-dotted line indicates the colour-magnitude relation measured by \citet{mieske10} for the metal-poor GCs and the solid and dashed lines show the mass-metallicity relations translated to colour-magnitude space (see the text). The good agreement with the slope of the dash-dotted line suggests that the blue tilt reflects the slope of the galaxy mass-metallicity relation of metal-poor proto-GC hosts, i.e.\ at $10^5\la\ms/\msun\la10^7$ and $z\ga2$.
}
\end{figure*}
At fixed redshift, \autoref{fig:f2} shows that the maximum cluster mass is a steep function of the metallicity. This translates the galaxy mass-metallicity relation into a maximum initial cluster mass-metallicity relation, which is visualised at different redshifts in \autoref{fig:f3} (left). The slope and normalisation of the lines are reminiscent of the mass-metallicity relation observed for metal-poor GCs in nearby galaxies \citep[the `blue tilt', e.g.][]{harris06,mieske10}. Previously, this relation had been attributed to non-standard physics, such as self-enrichment through multiple generations of star formation in massive GCs \citep{bailin09} or a bottom-heavy stellar initial mass function \citep{goudfrooij14}. More recently, explanations based on known physics have been put forward, such as the limited gas reservoirs available at low galaxy masses \citep{choksi18} or a low maximum mass scale for gravitational collapse \citep{usher18}, assuming that cluster formation proceeds in hydrostatic equilibrium discs \citep{reinacampos17}. The translation of the galaxy mass-metallicity relation into a maximum GC mass-metallicity relation falls in the latter category of known physics (and does not exclude the interpretations by \citealt{choksi18} and \citealt{usher18}, the latter of which likely sets the maximum GC mass at $\feh\ga-1.8$), but differs by not requiring any ingredients beyond an empirical galactic scaling relation.

To quantify whether the galaxy mass-metallicity relation may indeed explain the observed blue tilt, the maximum initial cluster mass-metallicity relation needs to be translated into a GC colour-magnitude relation. To convert metallicities to colours, the colour-metallicity relation from \citet[eq.~B1]{usher12} is used, whereas the masses are converted to magnitudes by assuming the GCs lost two thirds of their initial mass by stellar evolution and disruption \citep{reinacampos18} and adopting a $z$-band mass-to-light ratio of $M/L_z=2~\msun~{\rm L}_\odot^{-1}$. The result is shown in \autoref{fig:f3} (right), which compares the predicted relation between the minimum $z$-band magnitude and the $(g-z)$ colour to the observed colour-magnitude distribution of GCs in the Virgo \citep{jordan09} and Fornax \citep{jordan15} Clusters \citep[similar results have been obtained for other filters, see e.g.][]{usher18}. The slope of the predicted relation is highly similar to the observed slope of the blue tilt (dash-dotted line in \autoref{fig:f3}). This quantitatively supports the idea that the blue tilt reflects the slope of the galaxy mass-metallicity relation of metal-poor proto-GC host galaxies, i.e.\ at $10^5\la\ms/\msun\la10^7$ and $z\ga2$. In this interpretation, variations of the blue tilt slope between or within galaxies may reflect variations in the formation redshifts of GC (sub-)populations.

\begin{figure}
\includegraphics[width=\hsize]{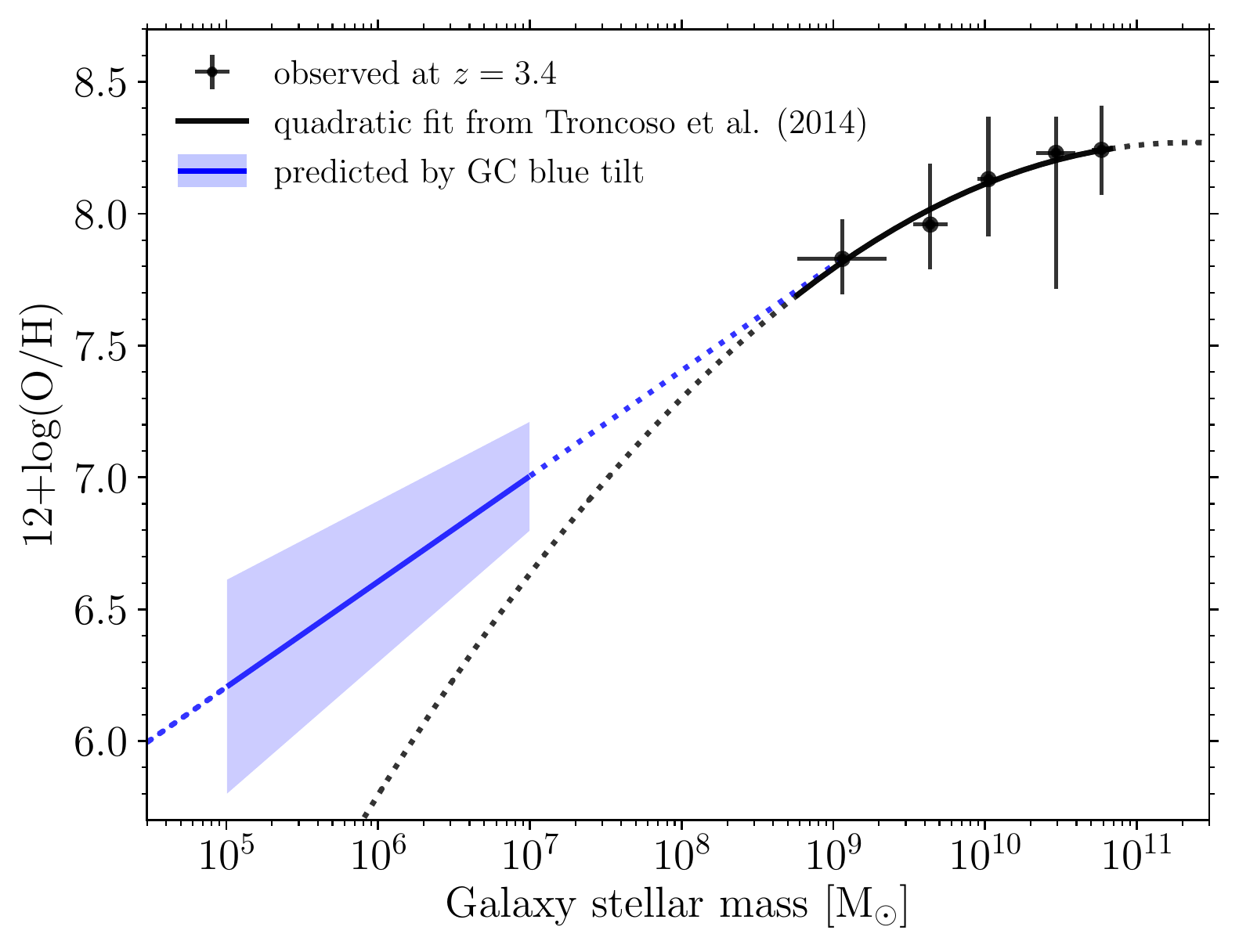}%
\caption{
\label{fig:f4}
Predicted slope of the gas phase mass-metallicity relation for the host galaxies of metal-poor proto-GCs, i.e.\ galaxies with $10^5\la\ms/\msun\la10^7$ and $z\ga2$, using the blue tilt of GCs at $z=0$ as in equation~(\ref{eq:infmmr}). The result is compared to existing measurements of the mass-metallicity relation at $z\sim3.4$ from \citet{troncoso14}. The predicted slope is similar to that at $\ms\sim10^9~\msun$, suggesting that $\oh \propto \ms^{0.4\pm0.1}$ for $10^5\la\ms/\msun\la10^9$. Future observations will be able to test this prediction.
}
\end{figure}

If the above interpretation of \autoref{fig:f3} is correct and $\Gamma$ does not change strongly with $\feh$, then the slope of the blue tilt directly traces the slope of the galaxy mass-metallicity relation. The conversion factor depends exclusively on the adopted colour-metallicity relation and the relation between the mass-to-light ratio and cluster mass and metallicity. Writing $\feh \propto \ms^\alpha$, $\feh=A(g-z)+B$, and adopting a constant mass-to-light ratio as above, we find
\be
\label{eq:infmmr}
\alpha = -2.5A\times\frac{{\rm d}(g-z)}{{\rm d}M_z} \approx (-13.7\pm2.4) \times \frac{{\rm d}(g-z)}{{\rm d}M_z} .
\ee
for $10^5\la\ms/\msun\la10^7$ and $z\ga2$, where the second equality again uses the colour-metallicity relation from \citet{usher12} and includes the associated uncertainty. Observations of the GC population in the Virgo and Fornax Clusters show that ${\rm d}(g-z)/{\rm d}M_z=-0.0293\pm0.0085$ \citep{mieske10}, which results in $\alpha=0.4\pm0.1$. This is a prediction for future observations of the mass-metallicity relation of ULLGs at $z>2$. \autoref{fig:f4} visualises this by showing how the predicted mass-metallicity relation of ULLGs connects to the observed gas phase mass-metallicity relation at $z\sim3.4$ \citep{troncoso14}. The slope matches that at $\ms\sim10^9~\msun$, suggesting that it may be constant at $\alpha=0.4\pm0.1$ for $10^5\la\ms/\msun\la10^9$ and $z>2$, consistent with \citet{ma16}. \looseness=-1

One caveat is in order when applying the above conversion. The slope of the minimum $z$-band magnitude of metal-poor GCs as a function of their $(g-z)$ colour may not be trivially identical to that of the expected blue tilt. After all, only the upper envelope of the GC $M_z{-}(g-z)$ distribution is truncated by limiting the initial cluster mass as a function of metallicity. As a result, the median colour at each magnitude may change more slowly than the truncation does. Given the assumptions necessary to reach the comparison made in \autoref{fig:f3}, a quantitative discussion of this effect is beyond the scope of this work. Future, more detailed studies of GC formation during galaxy formation and evolution \citep[e.g.\ along the lines of][]{pfeffer18} will be able to quantify more accurately how the galaxy mass-metallicity relation is related to the blue tilt.

\section{Proto-GCs host galaxies at high redshift}
\begin{figure}
\includegraphics[width=\hsize]{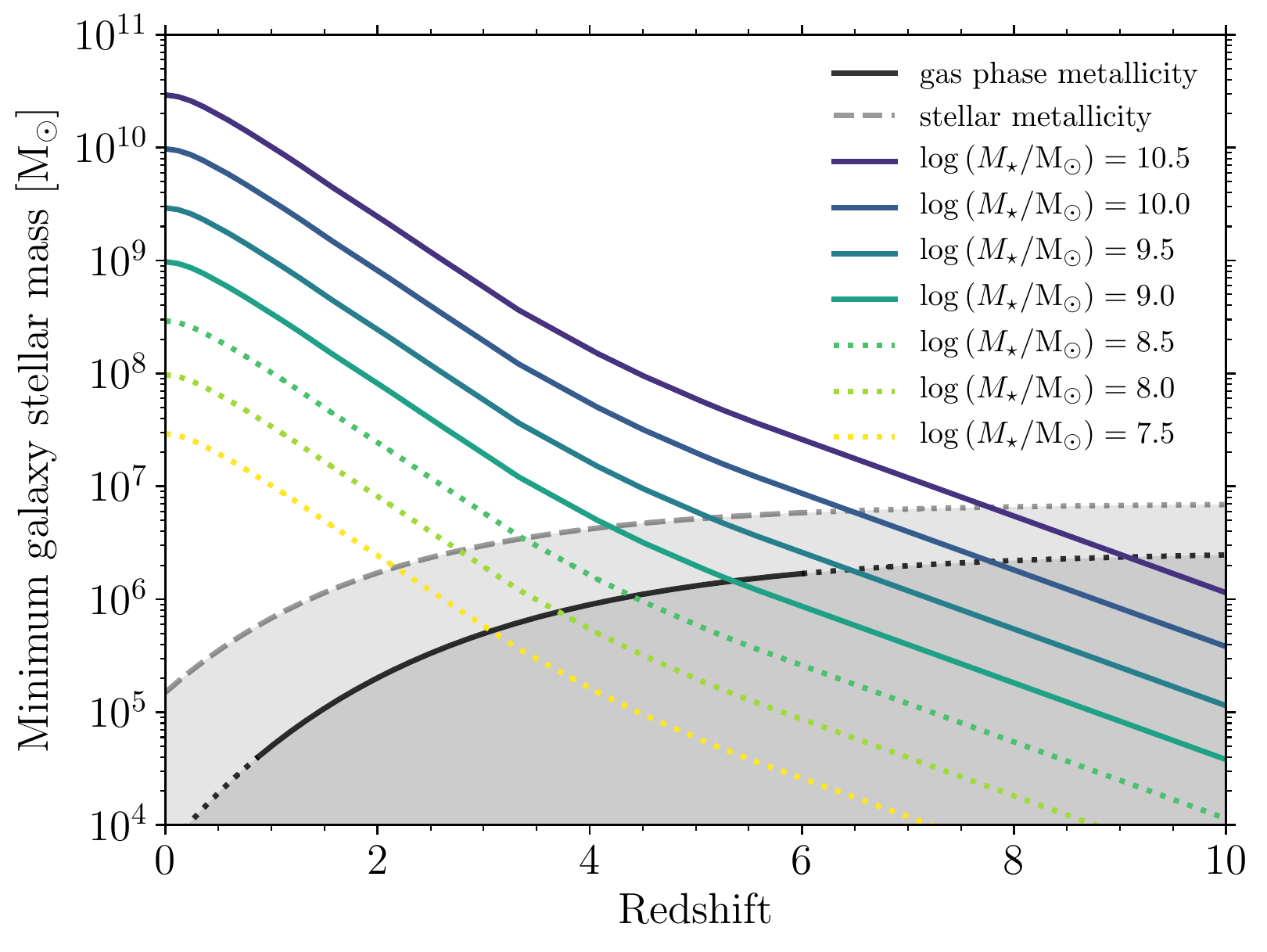}%
\caption{
\label{fig:f5}
Minimum galaxy mass for forming proto-GCs with $\feh\geq-2.5$ (grey lines) as a function of redshift. Coloured lines show the stellar mass growth histories of galaxies with $z=0$ masses in the range $3\times10^7\leq\ms/\msun\leq3\times10^{10}$ predicted by \citet[see the text]{behroozi13}. The grey shading marks a `zone of avoidance', where GC formation is rare.
}
\end{figure}

Using the empirical result that few GCs exist with $\feh<-2.5$, the mass-metallicity-redshift relations from \autoref{fig:f1} provide the typical minimum galaxy mass needed to form GCs as a function of redshift. This traces the intersection of the lines with the grey-shaded area in \autoref{fig:f1} and does not depend on the physical origin of the GC metallicity floor. \autoref{fig:f5} shows the result as grey lines. Because galaxy metallicities increase with time, galaxies forming GCs with $\feh=-2.5$ have lower masses at lower redshifts.

In order to determine the properties of the galaxy in which observed proto-GCs may end up, the host galaxy mass at the time of (proto-)GC formation can be connected to the average $z=0$ mass using a simple prescription for galaxy growth obtained by combining abundance matching and empirical constraints. \citet[fig.\ 12]{behroozi13} provide the fraction of the final stellar mass assembled as a function of redshift and halo mass at $z=0$, showing that it is very similar for haloes of $10^{11}\leq\mh/\msun\leq10^{12}$, with differences $<0.2~\dex$ at any redshift. In the model, the corresponding $z=0$ stellar mass range is $8\times10^8\leq\ms/\msun\leq4\times10^{10}$. The following adopts the relative stellar mass growth history of a halo with $\mh(z=0)=10^{11}~\msun$ for $3\times10^7\leq\ms/\msun\leq3\times10^{10}$, which thus represents an extrapolation for $\ms<8\times10^8~\msun$.

The coloured lines in \autoref{fig:f5} show the average stellar mass growth histories obtained following the above representation of the \citet{behroozi13} model for $z=0$ stellar masses in the range $3\times10^7\leq\ms/\msun\leq3\times10^{10}$. A comparison to the instantaneous minimum galaxy mass for forming proto-GCs with $\feh=-2.5$ (grey lines) quantifies the earlier statement that there exists a minimum galaxy mass and a maximum redshift for (proto-)GC formation. Any observed proto-GCs at $z\geq9$ require chemical enrichment to have proceeded so rapidly, that they are most likely to be forming in haloes that end up being more massive than the Milky Way at $z=0$, with $\mh\geq10^{12}~\msun$ and $\ms\geq3\times10^{10}~\msun$.

\begin{figure}
\includegraphics[width=\hsize]{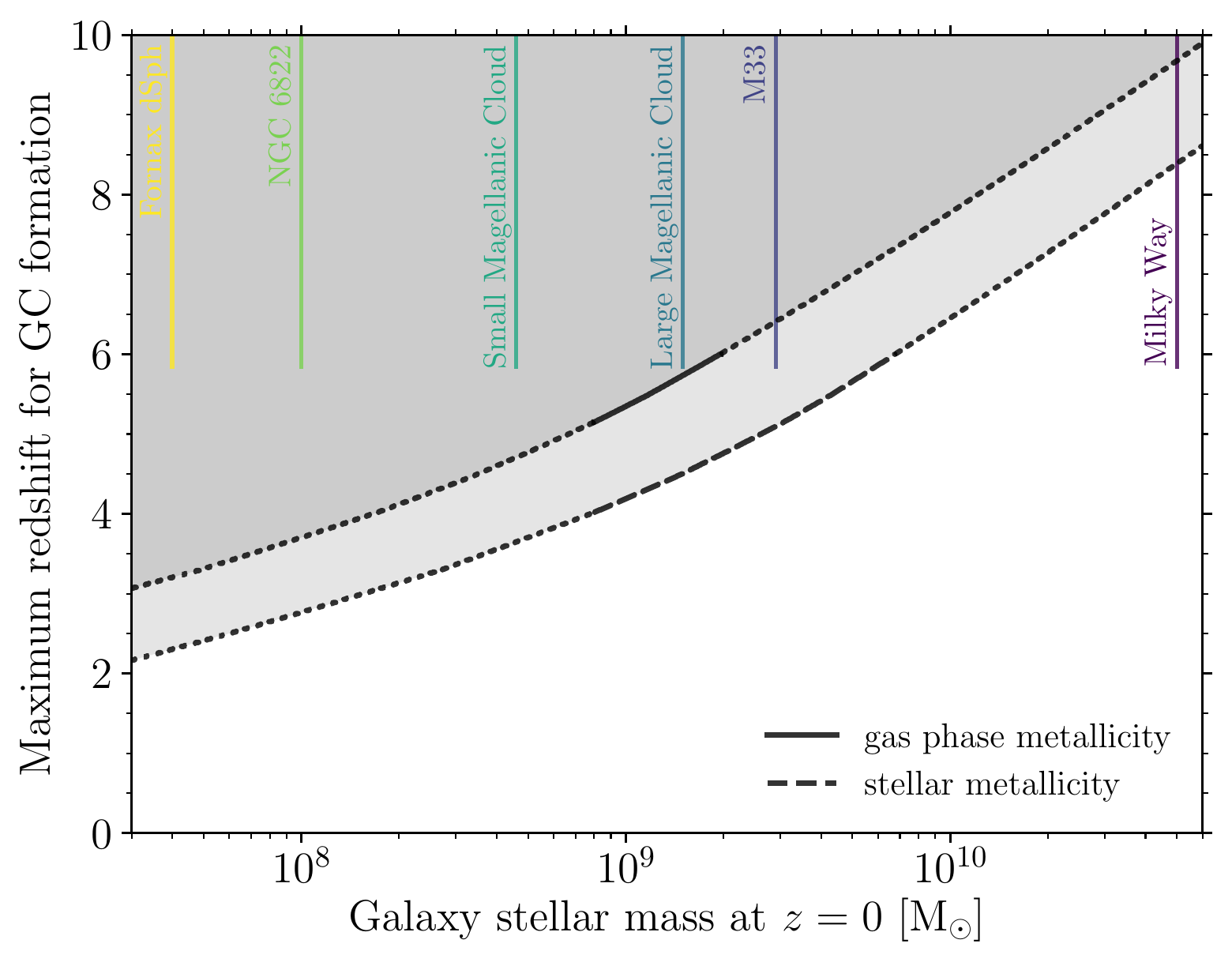}%
\caption{
\label{fig:f6}
Maximum redshift for the formation of in-situ proto-GCs (i.e.\ with $\feh\geq-2.5$) as a function of the $z=0$ galaxy stellar mass. The grey shading marks a `zone of avoidance', where on average GC formation is not expected to have taken place. Vertical lines mark example galaxies.
}
\end{figure}

\autoref{fig:f5} predicts a maximum redshift for GC formation as a function of the present-day galaxy mass, traced by the intersections of the coloured and grey lines. This is visualised explicitly in \autoref{fig:f6}, which shows the maximum formation redshift of in-situ proto-GCs (again defined by having $\feh\geq-2.5$) as a function of the $z=0$ galaxy stellar mass. The figure confirms that GCs in Milky Way-mass galaxies on average have a maximum formation redshift of $z\sim9$, with any earlier GC formation mostly taking place in much rarer, more massive haloes. This expectation is remarkably consistent with detailed simulations of the GC formation history in galaxies of Milky Way masses and lower, which predict an absence of GC formation at $z>7.5$ \citep{reinacampos19} or $z>9$ \citep{choksi18}. On the low-mass end of the mass range, galaxies like the Fornax dwarf spheroidal galaxy ($\ms\approx4\times10^7~\msun$) have a maximum GC formation redshift of $z\sim3$. Within uncertainties, this is consistent with GC ages in Fornax \citep{deboer16}, but deviations are expected due to stochastic low-mass galaxy growth (see Section~\ref{sec:concl}). For instance, the stellar mass of Fornax likely grew more rapidly than in \autoref{fig:f5} \citep{weisz14}. \looseness=-1

\section{Conclusion} \label{sec:concl}
In this Letter, the galaxy mass-metallicity relation is used to explain the observed minimum metallicity of GCs around $\feh\approx-2.5$, to infer its implications for the origin of the `blue tilt' and the mass-metallicity relation of ULLGs ($10^5\la\ms/\msun\la10^7$) at $z\ga2$, and to place limits on the demographics of the proto-GC population, to be observed with current and future facilities (e.g.\ JWST) using gravitational lensing. The conclusions of this work are as follows.
\begin{enumerate}
\item
For the galaxy mass-metallicity relation predicted by the FIRE simulations \citep{ma16}, the minimum GC metallicity corresponds to galaxy masses at which the total stellar mass is similar to that of GCs ($10^5{-}10^6~\msun$). For a fixed metallicity, the galaxy mass at a given redshift thus sets the maximum mass attainable by clusters forming at that redshift. (\autoref{fig:f1})
\item
An initial mass of $\mi\ga10^5~\msun$ is thought to be required for proto-GCs to survive for a Hubble time. Its similarity to the mean galaxy mass predicted by the mass-metallicity relation at $\feh=-2.5$ implies that the GC metallicity floor reflects the minimum galaxy mass needed to form long-lived GCs. (\autoref{fig:f2})
\item
Due to the close connection between the maximum cluster mass and the host galaxy mass, the maximum GC mass increases with metallicity for $\feh\la-1.8$. This explains the `blue tilt', i.e.\ the mass-metallicity and colour-magnitude relation of metal-poor GCs, as a natural result of the galaxy mass-metallicity relation. In this interpretation, the slope of the blue tilt traces the slope of the galaxy mass-metallicity relation at $z\ga2$, as described by equation~(\ref{eq:infmmr}). The observed blue tilt of GCs in the Virgo and Fornax Clusters implies a mass-metallicity relation slope of $\alpha=0.4\pm0.1$ at $z\ga2$ and $10^5\la\ms/\msun\la10^7$, which is a key prediction for upcoming surveys of ULLGs at $z>2$. (\autoref{fig:f3} and \autoref{fig:f4})
\item
The GC metallicity floor at $\feh\approx-2.5$ implies the existence of a minimum galaxy mass and a maximum redshift for GC formation. Using abundance matching models for galaxy growth to connect these masses and redshifts to galaxy masses at $z=0$, it is found that the maximum redshift for proto-GC formation increases as a function of the present-day galaxy mass. Any proto-GCs forming at $z>9$ are most likely to end up in galaxies presently more massive than the Milky Way, whereas GCs in low-mass galaxies such as the Fornax dwarf spheroidal ($\ms\approx4\times10^7~\msun$) on average formed at $z\la3$. (\autoref{fig:f5} and \autoref{fig:f6})
\item
In the context of galactic archaeological studies of the GC population, the above results imply that the first GCs that formed ended up in the most massive haloes and represent the in-situ population. On average, ex-situ GCs brought in by satellite galaxies formed later due to the lower masses of their host galaxies \citep[cf.][]{leaman13,kruijssen18c}. Testing these predictions using local-Universe age dating requires a precision of $\Delta\tau\sim0.2~\gyr$ for a redshift precision of $\Delta z\sim1$ at $z=6$. This is currently out of reach, but may be achieved with JWST \citep{kalirai18}.
\end{enumerate}
The above results represent statistical guidelines on the scale of the galaxy population. However, individual galaxies will deviate from these expectations, among others due to scatter on the adopted galactic scaling relations. For instance, if a low-mass galaxy assembled unusually rapidly, it likely formed proto-GCs in the zone of avoidance of \autoref{fig:f6}. With JWST, the incidence of these deviations from the ensemble average could plausibly be used to constrain the stochasticity of early galaxy formation at high redshift.

\section*{Acknowledgements}
JMDK is supported by the German Research Foundation (DFG; KR4801/1-1), the European Research Council (ERC; ERC-StG-714907), and SFB 881 (subproject B2) of the DFG. The participants of the Lorentz Center workshop `Formation of Stars and Massive clusters in Dwarf Galaxies over Cosmic Time' (February 18--22, 2019) and particularly the SOC chair, Rychard Bouwens, are thanked for inspiring conversations that led to the development of this work. N.~Bastian, I.~Cabrera-Ziri, M.~Chevance, N.~Choksi, M.~Gieles, O.~Gnedin, S.~Larsen, M.~Maseda, J.~Pfeffer, M.~Reina-Campos, A.~Sills, S.~Trujillo-Gomez, C.~Usher, E.~Vanzella, and an anonymous referee are thanked for helpful feedback.
\vspace{-4mm}

\bibliographystyle{mnras}
\bibliography{mybib}

\begin{thebibliography}{}
\makeatletter
\relax
\def\mn@urlcharsother{\let\do\@makeother \do\$\do\&\do\#\do\^\do\_\do\%\do\~}
\def\mn@doi{\begingroup\mn@urlcharsother \@ifnextchar [ {\mn@doi@}
  {\mn@doi@[]}}
\def\mn@doi@[#1]#2{\def\@tempa{#1}\ifx\@tempa\@empty \href
  {http://dx.doi.org/#2} {doi:#2}\else \href {http://dx.doi.org/#2} {#1}\fi
  \endgroup}
\def\mn@eprint#1#2{\mn@eprint@#1:#2::\@nil}
\def\mn@eprint@arXiv#1{\href {http://arxiv.org/abs/#1} {{\tt arXiv:#1}}}
\def\mn@eprint@dblp#1{\href {http://dblp.uni-trier.de/rec/bibtex/#1.xml}
  {dblp:#1}}
\def\mn@eprint@#1:#2:#3:#4\@nil{\def\@tempa {#1}\def\@tempb {#2}\def\@tempc
  {#3}\ifx \@tempc \@empty \let \@tempc \@tempb \let \@tempb \@tempa \fi \ifx
  \@tempb \@empty \def\@tempb {arXiv}\fi \@ifundefined
  {mn@eprint@\@tempb}{\@tempb:\@tempc}{\expandafter \expandafter \csname
  mn@eprint@\@tempb\endcsname \expandafter{\@tempc}}}

\bibitem[\protect\citeauthoryear{{Adamo}, {Kruijssen}, {Bastian}, {Silva-Villa}
   \& {Ryon}}{{Adamo} et~al.}{2015}]{adamo15b}
{Adamo} A.,  {Kruijssen} J.~M.~D.,  {Bastian} N.,  {Silva-Villa} E.,   {Ryon}
  J.,  2015, \mn@doi [\mnras] {10.1093/mnras/stv1203}, \href
  {http://adsabs.harvard.edu/abs/2015MNRAS.452..246A} {452, 246}

\bibitem[\protect\citeauthoryear{{Bailin} \& {Harris}}{{Bailin} \&
  {Harris}}{2009}]{bailin09}
{Bailin} J.,  {Harris} W.~E.,  2009, \mn@doi [\apj]
  {10.1088/0004-637X/695/2/1082}, \href
  {http://adsabs.harvard.edu/abs/2009ApJ...695.1082B} {695, 1082}

\bibitem[\protect\citeauthoryear{{Bastian}}{{Bastian}}{2008}]{bastian08}
{Bastian} N.,  2008, \mn@doi [\mnras] {10.1111/j.1365-2966.2008.13775.x}, \href
  {http://adsabs.harvard.edu/abs/2008MNRAS.390..759B} {390, 759}

\bibitem[\protect\citeauthoryear{{Beasley}, {Leaman}, {Gallart}, {Larsen},
  {Battaglia}, {Monelli}  \& {Pedreros}}{{Beasley} et~al.}{2019}]{beasley19}
{Beasley} M.~A.,  et~al.,  2019, \mnras~submitted, arXiv:1904.01084

\bibitem[\protect\citeauthoryear{{Behroozi}, {Wechsler}  \&
  {Conroy}}{{Behroozi} et~al.}{2013}]{behroozi13}
{Behroozi} P.~S.,  {Wechsler} R.~H.,   {Conroy} C.,  2013, \mn@doi [\apj]
  {10.1088/0004-637X/770/1/57}, \href
  {http://adsabs.harvard.edu/abs/2013ApJ...770...57B} {770, 57}

\bibitem[\protect\citeauthoryear{{Bouwens}, {Illingworth}, {Oesch}, {Maseda},
  {Ribeiro}, {Stefanon}  \& {Lam}}{{Bouwens} et~al.}{2017}]{bouwens17}
{Bouwens} R.~J.,  et~al.,  2017, arXiv:1711.02090

\bibitem[\protect\citeauthoryear{{Boylan-Kolchin}}{{Boylan-Kolchin}}{2018}]{boylankolchin18}
{Boylan-Kolchin} M.,  2018, \mn@doi [\mnras] {10.1093/mnras/sty1490}, \href
  {http://adsabs.harvard.edu/abs/2018MNRAS.479..332B} {479, 332}

\bibitem[\protect\citeauthoryear{{Choksi}, {Gnedin}  \& {Li}}{{Choksi}
  et~al.}{2018}]{choksi18}
{Choksi} N.,  {Gnedin} O.~Y.,   {Li} H.,  2018, \mn@doi [\mnras]
  {10.1093/mnras/sty1952}, \href
  {http://adsabs.harvard.edu/abs/2018MNRAS.480.2343C} {480, 2343}

\bibitem[\protect\citeauthoryear{{Forbes} et~al.,}{{Forbes}
  et~al.}{2018}]{forbes18}
{Forbes} D.~A.,  et~al., 2018, \mn@doi [Proc. R. Soc. A]
  {10.1098/rspa.2017.0616}, \href
  {http://adsabs.harvard.edu/abs/2018RSPSA.47470616F} {474, 20170616}

\bibitem[\protect\citeauthoryear{{Goudfrooij} \& {Kruijssen}}{{Goudfrooij} \&
  {Kruijssen}}{2014}]{goudfrooij14}
{Goudfrooij} P.,  {Kruijssen} J.~M.~D.,  2014, \mn@doi [\apj]
  {10.1088/0004-637X/780/1/43}, \href
  {http://adsabs.harvard.edu/abs/2014ApJ...780...43G} {780, 43}

\bibitem[\protect\citeauthoryear{{Harris}, {Whitmore}, {Karakla}, {Oko{\'n}},
  {Baum}, {Hanes}  \& {Kavelaars}}{{Harris} et~al.}{2006}]{harris06}
{Harris} W.~E.,  et~al.,  2006, \mn@doi [\apj]
  {10.1086/498058}, \href {http://adsabs.harvard.edu/abs/2006ApJ...636...90H}
  {636, 90}

\bibitem[\protect\citeauthoryear{{Hopkins} et~al.,}{{Hopkins}
  et~al.}{2018}]{hopkins18}
{Hopkins} P.~F.,  et~al., 2018, \mn@doi [\mnras] {10.1093/mnras/sty1690}, \href
  {http://adsabs.harvard.edu/abs/2018MNRAS.480..800H} {480, 800}

\bibitem[\protect\citeauthoryear{{Johnson} et~al.,}{{Johnson}
  et~al.}{2017}]{johnson17}
{Johnson} T.~L.,  et~al., 2017, \mn@doi [\apj] {10.3847/1538-4357/aa7756},
  \href {http://adsabs.harvard.edu/abs/2017ApJ...843...78J} {843, 78}

\bibitem[\protect\citeauthoryear{{Jord{\'a}n} et~al.,}{{Jord{\'a}n}
  et~al.}{2007}]{jordan07}
{Jord{\'a}n} A.,  et~al., 2007, \mn@doi [\apjs] {10.1086/516840}, \href
  {http://adsabs.harvard.edu/abs/2007ApJS..171..101J} {171, 101}

\bibitem[\protect\citeauthoryear{{Jord{\'a}n} et~al.,}{{Jord{\'a}n}
  et~al.}{2009}]{jordan09}
{Jord{\'a}n} A.,  et~al., 2009, \mn@doi [\apjs] {10.1088/0067-0049/180/1/54},
  \href {http://adsabs.harvard.edu/abs/2009ApJS..180...54J} {180, 54}

\bibitem[\protect\citeauthoryear{{Jord{\'a}n}, {Peng}, {Blakeslee},
  {C{\^o}t{\'e}}, {Eyheramendy}  \& {Ferrarese}}{{Jord{\'a}n}
  et~al.}{2015}]{jordan15}
{Jord{\'a}n} A.,  {Peng} E.~W.,  {Blakeslee} J.~P.,  {C{\^o}t{\'e}} P.,
  {Eyheramendy} S.,   {Ferrarese} L.,  2015, \mn@doi [\apjs]
  {10.1088/0067-0049/221/1/13}, \href
  {http://adsabs.harvard.edu/abs/2015ApJS..221...13J} {221, 13}

\bibitem[\protect\citeauthoryear{{Kalirai}}{{Kalirai}}{2018}]{kalirai18}
{Kalirai} J.,  2018, \mn@doi [Contemporary Physics]
  {10.1080/00107514.2018.1467648}, \href
  {http://adsabs.harvard.edu/abs/2018ConPh..59..251K} {59, 251}

\bibitem[\protect\citeauthoryear{{Kruijssen}}{{Kruijssen}}{2012}]{kruijssen12d}
{Kruijssen} J.~M.~D.,  2012, \mn@doi [\mnras]
  {10.1111/j.1365-2966.2012.21923.x}, \href
  {http://adsabs.harvard.edu/abs/2012MNRAS.426.3008K} {426, 3008}

\bibitem[\protect\citeauthoryear{{Kruijssen}}{{Kruijssen}}{2015}]{kruijssen15b}
{Kruijssen} J.~M.~D.,  2015, \mn@doi [\mnras] {10.1093/mnras/stv2026}, \href
  {http://adsabs.harvard.edu/abs/2015MNRAS.454.1658K} {454, 1658}

\bibitem[\protect\citeauthoryear{{Kruijssen}, {Pfeffer}, {Reina-Campos},
  {Crain}  \& {Bastian}}{{Kruijssen} et~al.}{2018}]{kruijssen18c}
{Kruijssen} J.~M.~D.,  {Pfeffer} J.~L.,  {Reina-Campos} M.,  {Crain} R.~A.,
  {Bastian} N.,  2018, \mnras~in press, arXiv:1806.05680

\bibitem[\protect\citeauthoryear{{Leaman}, {VandenBerg}  \& {Mendel}}{{Leaman}
  et~al.}{2013}]{leaman13}
{Leaman} R.,  {VandenBerg} D.~A.,   {Mendel} J.~T.,  2013, \mn@doi [\mnras]
  {10.1093/mnras/stt1540}, \href
  {http://adsabs.harvard.edu/abs/2013MNRAS.436..122L} {436, 122}

\bibitem[\protect\citeauthoryear{{Ma}, {Hopkins}, {Faucher-Gigu{\`e}re},
  {Zolman}, {Muratov}, {Kere{\v s}}  \& {Quataert}}{{Ma} et~al.}{2016}]{ma16}
{Ma} X.,  et~al.,  2016, \mn@doi [\mnras]
  {10.1093/mnras/stv2659}, \href
  {http://adsabs.harvard.edu/abs/2016MNRAS.456.2140M} {456, 2140}

\bibitem[\protect\citeauthoryear{{Maiolino} \& {Mannucci}}{{Maiolino} \&
  {Mannucci}}{2019}]{maiolino19}
{Maiolino} R.,  {Mannucci} F.,  2019, \mn@doi [\aapr]
  {10.1007/s00159-018-0112-2}, \href
  {http://adsabs.harvard.edu/abs/2019A%26ARv..27....3M} {27, 3}

\bibitem[\protect\citeauthoryear{{Mieske} et~al.,}{{Mieske}
  et~al.}{2010}]{mieske10}
{Mieske} S.,  et~al., 2010, \mn@doi [\apj] {10.1088/0004-637X/710/2/1672},
  \href {http://adsabs.harvard.edu/abs/2010ApJ...710.1672M} {710, 1672}

\bibitem[\protect\citeauthoryear{{Pfeffer}, {Kruijssen}, {Crain}  \&
  {Bastian}}{{Pfeffer} et~al.}{2018}]{pfeffer18}
{Pfeffer} J.,  {Kruijssen} J.~M.~D.,  {Crain} R.~A.,   {Bastian} N.,  2018,
  \mn@doi [\mnras] {10.1093/mnras/stx3124}, \href
  {http://adsabs.harvard.edu/abs/2018MNRAS.475.4309P} {475, 4309}

\bibitem[\protect\citeauthoryear{{Pfeffer}, {Bastian}, {Crain}, {Kruijssen},
  {Hughes}  \& {Reina-Campos}}{{Pfeffer} et~al.}{2019}]{pfeffer19}
{Pfeffer} J.,  {Bastian} N.,  {Crain} R.~A.,  {Kruijssen} J.~M.~D.,  {Hughes}
  M.~E.,   {Reina-Campos} M.,  2019, \mnras~submitted

\bibitem[\protect\citeauthoryear{{Reina-Campos} \& {Kruijssen}}{{Reina-Campos}
  \& {Kruijssen}}{2017}]{reinacampos17}
{Reina-Campos} M.,  {Kruijssen} J.~M.~D.,  2017, \mn@doi [\mnras]
  {10.1093/mnras/stx790}, \href
  {http://adsabs.harvard.edu/abs/2017MNRAS.469.1282R} {469, 1282}

\bibitem[\protect\citeauthoryear{{Reina-Campos}, {Kruijssen}, {Pfeffer},
  {Bastian}  \& {Crain}}{{Reina-Campos} et~al.}{2018}]{reinacampos18}
{Reina-Campos} M.,  {Kruijssen} J.~M.~D.,  {Pfeffer} J.,  {Bastian} N.,
  {Crain} R.~A.,  2018, \mn@doi [\mnras] {10.1093/mnras/sty2451}, \href
  {http://adsabs.harvard.edu/abs/2018MNRAS.481.2851R} {481, 2851}

\bibitem[\protect\citeauthoryear{{Reina-Campos}, {Kruijssen}, {Pfeffer},
  {Bastian}  \& {Crain}}{{Reina-Campos} et~al.}{2019}]{reinacampos19}
{Reina-Campos} M.,  {Kruijssen} J.~M.~D.,  {Pfeffer} J.,  {Bastian} N.,
  {Crain} R.~A.,  2019, \mnras~submitted

\bibitem[\protect\citeauthoryear{{Renzini}}{{Renzini}}{2017}]{renzini17}
{Renzini} A.,  2017, \mn@doi [\mnras] {10.1093/mnrasl/slx057}, \href
  {http://adsabs.harvard.edu/abs/2017MNRAS.469L..63R} {469, L63}

\bibitem[\protect\citeauthoryear{{Tacconi} et~al.,}{{Tacconi}
  et~al.}{2018}]{tacconi18}
{Tacconi} L.~J.,  et~al., 2018, \mn@doi [\apj] {10.3847/1538-4357/aaa4b4},
  \href {http://adsabs.harvard.edu/abs/2018ApJ...853..179T} {853, 179}

\bibitem[\protect\citeauthoryear{{Tolstoy}, {Hill}  \& {Tosi}}{{Tolstoy}
  et~al.}{2009}]{tolstoy09}
{Tolstoy} E.,  {Hill} V.,   {Tosi} M.,  2009, \mn@doi [\araa]
  {10.1146/annurev-astro-082708-101650}, \href
  {http://adsabs.harvard.edu/abs/2009ARA%26A..47..371T} {47, 371}

\bibitem[\protect\citeauthoryear{{Troncoso} et~al.,}{{Troncoso}
  et~al.}{2014}]{troncoso14}
{Troncoso} P.,  et~al., 2014, \mn@doi [\aap] {10.1051/0004-6361/201322099},
  \href {http://ads.ari.uni-heidelberg.de/abs/2014A%26A...563A..58T} {563, A58}

\bibitem[\protect\citeauthoryear{{Usher} et~al.,}{{Usher}
  et~al.}{2012}]{usher12}
{Usher} C.,  et~al., 2012, \mn@doi [\mnras] {10.1111/j.1365-2966.2012.21801.x},
  \href {http://adsabs.harvard.edu/abs/2012MNRAS.426.1475U} {426, 1475}

\bibitem[\protect\citeauthoryear{{Usher}, {Pfeffer}, {Bastian}, {Kruijssen},
  {Crain}  \& {Reina-Campos}}{{Usher} et~al.}{2018}]{usher18}
{Usher} C.,  {Pfeffer} J.,  {Bastian} N.,  {Kruijssen} J.~M.~D.,  {Crain}
  R.~A.,   {Reina-Campos} M.,  2018, \mn@doi [\mnras] {10.1093/mnras/sty1895},
  \href {http://adsabs.harvard.edu/abs/2018MNRAS.480.3279U} {480, 3279}

\bibitem[\protect\citeauthoryear{{Vanzella} et~al.,}{{Vanzella}
  et~al.}{2017}]{vanzella17}
{Vanzella} E.,  et~al., 2017, \mn@doi [\mnras] {10.1093/mnras/stx351}, \href
  {http://adsabs.harvard.edu/abs/2017MNRAS.467.4304V} {467, 4304}

\bibitem[\protect\citeauthoryear{{Vanzella} et~al.,}{{Vanzella}
  et~al.}{2019}]{vanzella19}
{Vanzella} E.,  et~al., 2019, \mn@doi [\mnras] {10.1093/mnras/sty3311}, \href
  {http://adsabs.harvard.edu/abs/2019MNRAS.483.3618V} {483, 3618}

\bibitem[\protect\citeauthoryear{{Weisz}, {Dolphin}, {Skillman}, {Holtzman},
  {Gilbert}, {Dalcanton}  \& {Williams}}{{Weisz} et~al.}{2014}]{weisz14}
{Weisz} D.~R.,  et~al.,  2014, \mn@doi [\apj]
  {10.1088/0004-637X/789/2/147}, \href
  {http://adsabs.harvard.edu/abs/2014ApJ...789..147W} {789, 147}

\bibitem[\protect\citeauthoryear{{Zick}, {Weisz}  \& {Boylan-Kolchin}}{{Zick}
  et~al.}{2018}]{zick18}
{Zick} T.~O.,  {Weisz} D.~R.,   {Boylan-Kolchin} M.,  2018, \mn@doi [\mnras]
  {10.1093/mnras/sty662}, \href
  {http://adsabs.harvard.edu/abs/2018MNRAS.477..480Z} {477, 480}

\bibitem[\protect\citeauthoryear{{de Boer} \& {Fraser}}{{de Boer} \&
  {Fraser}}{2016}]{deboer16}
{de Boer} T.~J.~L.,  {Fraser} M.,  2016, \mn@doi [\aap]
  {10.1051/0004-6361/201527580}, \href
  {http://adsabs.harvard.edu/abs/2016A%26A...590A..35D} {590, A35}

\makeatother
\end{thebibliography}

\vspace{-3mm}

\bsp

\label{lastpage}

\end{document}